\documentclass[aps,prd,preprint,showpacs,floats,nofootinbib]{revtex4-1}
\usepackage[dvips]{graphicx}
\usepackage{bm,bbm}
\usepackage{euscript,amsmath,amsfonts}

\begin{document}
\title{B physics constraints on a flavor symmetric scalar model to account for the $t{\bar t}$ asymmetry and Wjj excess at CDF}
\author {Guohuai Zhu}
\email[E-mail address: ]{zhugh@zju.edu.cn}
\affiliation{Zhejiang Institute of Modern Physics, Department of Physics, \\
 Zhejiang University, Hangzhou, Zhejiang 310027, P.R. China}

\date{\today}
\begin{abstract}
Recently Nelson {\it et al.} proposed an interesting flavor symmetric model to account for the top quark forward-backward asymmetry and
the dijet anomaly at CDF simultaneously with just three parameters: a coupling constant of order one, and two scalar masses of 160 GeV
and 220 GeV. However these fiducial values of the parameters lead to the branching ratio of a almost pure penguin $B \to \pi K$ decay
about one hundred times larger than the experimental results. Consider also the precision electroweak constraints, the scalar masses should
be at least around 500 GeV. Actually with the coupling constant larger than one, it is impossible to explain either of the two CDF measurements consistently in this model. But one may raise the charged scalar mass to, for example, $250$ GeV and reduce the coupling strength to $0.6$ to meet the B physics constraints. With this parameter set, the Wjj cross section is found to be in the right range. But due to the scalar mass splitting, its correction to T-parameter is about $3 \sigma$ away from the precision electroweak constraints. In addition, the top quark forward-backward asymmetry should be well below $0.1$ with this small coupling constant.
\end{abstract}

\maketitle

The CDF collaboration has recently updated the measurements on the forward-backward asymmetry in top quark pair production
with a larger data sample about $5.3$ fb$^{-1}$\cite{Aaltonen:2011kc,cdfdilepton}.
Interestingly, deviations from the Standard Model (SM) predictions are observed in the total forward-backward asymmetry both in the
semi-leptonic $t{\bar t}$ data and in the di-lepton channel. In addition, A distributional measurement found that
$A^{t\bar{t}}_{FB}(M_{t\bar{t}}> 450~\text{GeV})=0.475\pm 0.112$ in the $t\bar{t}$ rest frame, which deviates from the QCD correction
prediction $0.088 \pm 0.013$ by $3.5~\sigma$. The CDF collaboration has also reported another $3.2~\sigma$ anomaly in the $120-160$ GeV
range of the invariant dijet mass distribution in association with a W boson \cite{Aaltonen:2011mk}.

A flavor symmetric model was proposed in \cite{Nelson:2011us} to explain simultaneously the $t{\bar t}$ asymmetry and Wjj excess at CDF
\footnote{An alternative attempt can be found in \cite{Buckley:2011vc} by introducing a light leptophobic $Z^\prime$ gauge boson, though the predicted total cross section of $t\bar{t}$ production seems to be too small at the Tevatron (see, e.g., \cite{Shu:2011au})}.
A $\mathbb{Z}_3$ triplet of complex scalar fields $\Phi=(\Phi_1,\Phi_2,\Phi_3)$ is introduced in \cite{Nelson:2011us}.
These color-singlet weak-doublet scalars respect the flavor symmetry:
\begin{align}
\left ( \prod_{i=1}^3 U(1)_{q_{Li}} \times U(1)_{u_{Ri}} \right ) \times U(3)_{d_R} \times \mathbb{Z}_3~,
\end{align}
where $q_{Li}$ and $u_{Ri}$ have charge $+1$ under $U(1)_{q_{Li}}$ and $U(1)_{u_{Ri}}$, respectively, while $d_R$ is in a fundamental representation of $U(3)$. This flavor symmetry is also preserved in the SM without Yukawa interactions.

In this model the interaction of the scalars $\Phi$ with the SM quarks are completely determined by the flavor symmetry with a universal coupling strength. The $Wjj$ anomaly can then be interpreted as $u\bar{s} \to W^+ \Phi_3^0$ via a s-channel $\Phi_3^+$ exchange,
and $\Phi_3^0$ decays subsequently to a jet pair with its mass to be around $160$ GeV.
The top quark forward-backward asymmetry can be explained by $u\bar{u} \to t\bar{t}$ via a t-channel $\Phi^0_2$ exchange
and $d\bar{d} \to t\bar{t}$ via a t-channel $\Phi^+_2$ exchange. At first glance, this seems to be in contradiction with the observation of \cite{Shu:2009xf} that t-channel exchange of a color-singlet scalar has great difficulty to produce a large positive contribution to the top quark
forward-backward asymmetry. However a closer look at Fig. 2 of \cite{Shu:2009xf} reveals that there does have a narrow window with the scalar mass lighter than $250$ GeV.

However this flavor symmetry model also contributes to hadronic b decays. Although there is no new CP phase introduced,  we will show in the following that the effective operator $(\bar{b}_L u_R)(u_R s_L)$ via an exchange of such a light $\Phi$ is constrained severely by the penguin dominant processes, such as $B \to \pi K$ decays \footnote{The implications of rare B decays on t-channel models to account for the Tevatron
top-pair asymmetry have been discussed recently in \cite{Chen:2011mg}.  }.

In this flavor symmetry model, the color-singlet weak-doublet scalars $\Phi$ are charged $-1/2$ under $U(1)_Y$ and singlets under $U(3)_{d_R}$.
The interaction between $\Phi_i$ and the SM quarks \cite{Nelson:2011us}
\begin{align}\label{Eq:interaction}
 -\lambda (\bar{q}_{L1} \Phi_2 u_{R3} + \bar{q}_{L2} \Phi_3 u_{R1}+\bar{q}_{L3} \Phi_1 u_{R2}+ c.c.) 
\end{align}
is completely determined by the flavor symmetry in which $\Phi_i$ ($i=1,2,3$) are charged as
\begin{align}
\Phi_1 \sim (0,0,1)~, \hspace*{1cm} \Phi_2 \sim (1,0,0)~, \hspace*{1cm} \Phi_3 \sim (0,1,0)
\end{align}
under $U(1)_{q_{L1}} \times U(1)_{q_{L2}} \times U(1)_{q_{L3}}$, and charged as
\begin{align}
\Phi_1 \sim (0,-1,0)~, \hspace*{1cm} \Phi_2 \sim (0,0,-1)~, \hspace*{1cm} \Phi_3 \sim (-1,0,0)
\end{align}
under $U(1)_{u_{R1}} \times U(1)_{u_{R2}} \times U(1)_{u_{R3}}$. Then the only free parameters are the coupling constant $\lambda$ and the scalar masses
$m_{\Phi^0}$ and $m_{\Phi^-}$.

To interpret the CDF anomalies of Wjj and forward-backward asymmetry of top quark, $\lambda=1.4$, $m_{\Phi^0}=160$ GeV and $m_{\Phi^-}=220$ GeV have been chosen in \cite{Nelson:2011us} as "fiducial" values \footnote{The same-sign tops production is extremely suppressed in this model. Otherwise such light scalars might be severely constrained, see e.g. \cite{AguilarSaavedra:2011zy}. }. However in the mass basis, Eq. (\ref{Eq:interaction}) also generates effective four fermion operators,
among which contains
\begin{align}\label{Eq:bsuu}
{\cal H}_{eff}^\Phi=-\frac{\lambda^2}{m_{\Phi^-}^2} V_{cb}^\ast V_{cs} (\bar{b}_L u_R) (\bar{u}_R s_L)~.
\end{align}
As noticed in \cite{Nelson:2011us}, this operator contributes to the charmless process $b \to s\bar{u} u$ in comparison to the relevant effective Hamiltonian of the SM (where electroweak penguin operators have been neglected) \cite{Buchalla:1995vs}
\begin{align}
{\cal H}_{eff}=\frac{G_F}{\sqrt{2}} \left ( V_{ub}^\ast V_{us} \sum_{i=1}^6 C_i O_i + V_{cb}^\ast V_{cs} \sum_{i=3}^6 C_i O_i \right )~,
\end{align}
with
\begin{align}
O_1&=(\bar{b} u)_{V-A} (\bar{u} s)_{V-A} \hspace*{1.3cm} O_2=(\bar{b}_\alpha u_\beta)_{V-A} (\bar{u}_\beta s_\alpha)_{V-A}   \nonumber \\
O_3&=(\bar{b} s)_{V-A} (\bar{u} u)_{V-A} \hspace*{1.3cm} O_4=(\bar{b}_\alpha s_\beta)_{V-A} (\bar{u}_\beta u_\alpha)_{V-A} \nonumber \\
O_5&=(\bar{b} s)_{V-A} (\bar{u} u)_{V+A} \hspace*{1.3cm} O_6=-2\bar{b}(1+\gamma_5)u \bar{u} (1-\gamma_5) s
\end{align}
Since Eq. (\ref{Eq:bsuu}) is obtained at tree level, we will also consider the Wilson coefficients in the SM at leading order.
Matching the effective operators to the full theory at $\mu=M_W$, one finds $C_1(M_W)=1$ and other Wilson coefficients to be zero at leader order in the SM. But the flavor symmetry model contributes to $C_6$ as
\begin{align}
C_6^\Phi (M_W)&=\frac{\lambda^2}{8 m_{\Phi^-}^2} \left / \frac{G_F}{\sqrt{2}} \simeq 0.614 \right .
\end{align}
which is even comparable to $C_1(M_W)$ in the magnitude.

Running the scale down from $M_W$ to $m_b$, one finds in the SM
\begin{align}
C_1(m_b)&=1.115~,\hspace*{0.3cm} C_2(m_b)=-0.245~,\hspace*{0.3cm}C_3(m_b)=0.012~,\nonumber \\
C_4(m_b)&=-0.033~,\hspace*{0.3cm} C_5(m_b)=0.008~,\hspace*{0.3cm}C_6(m_b)=-0.038~.
\end{align}
But when the new scalar contributions are included, the Wilson coefficients $C_i$($i=3-6$) are changed to be
\begin{align}
C_3(m_b)&=0.062~,\hspace*{0.3cm}C_4(m_b)=-0.138~,\hspace*{0.3cm} C_5(m_b)=0.070~,\hspace*{0.3cm}C_6(m_b)=1.025~.
\end{align}
One may easily notice that $C_6(m_b)$ is surprisingly large in this flavor symmetry model. Even considering the theoretical uncertainties on hadronic B decays, it will lead to too large branching ratios on the penguin dominant decays, such as $B \to \pi K$ channels as we will show immediately.

For charmless B decays, there are three factorization approaches being widely used:
QCD factorization \cite{Beneke:1999br,Beneke:2000ry,Beneke:2003zv},
the perturbative QCD method (PQCD) \cite{Keum:2000ph,Keum:2000wi,Lu:2000em} and soft collinear effective theory (SCET)
\cite{Bauer:2000yr,Bauer:2001cu,Bauer:2004tj}. Here we will adopt QCD factorization method. Notice that the new physics amplitude is
calculated at tree level, correspondingly the Wilson coefficients are calculated at leading logarithm. To be consistent, the decay amplitudes of QCD factorization are also evaluated at leading order of $\alpha_s$. Let's consider the almost pure penguin process $B^+ \to \pi^+ K^0$ decay. Taking
$f_K=160$ MeV, the form factor $F^{B\pi}_0(0)=0.26$ \cite{Ball:2004ye,Duplancic:2008ix}, the current quark mass $m_s(2 \mbox{GeV})=100$ MeV \cite{Nakamura:2010zzi} and the relevant CKM parameters \cite{Charles:2004jd} $A=0.812$, $\lambda=0.2254$, we obtain
\begin{align}
{\cal B}(B^+ \to \pi^+ K^0)=2.4 \times 10^{-3}
\end{align}
which is about one hundred times larger than the experimental measurement $(23.1 \pm 1.0)\times 10^{-6}$ \cite{Nakamura:2010zzi}. Therefore the fiducial values of
$\lambda=1.4$, $m_{\Phi^-}=220$ GeV taken by Nelson {\it et al.} \cite{Nelson:2011us} are apparently inconsistent with the penguin dominant B decays.

\begin{figure}
\includegraphics[scale=1.0]{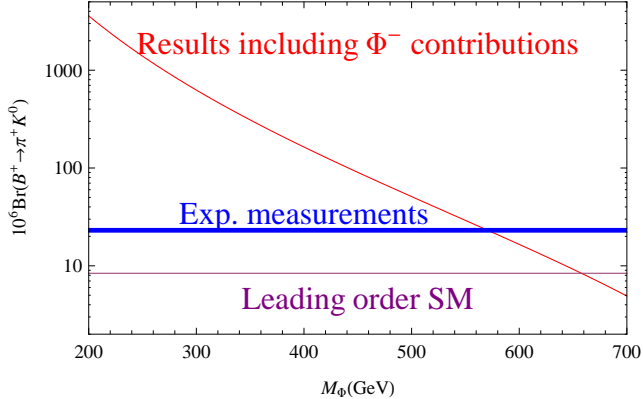} \caption{\label{fig:1} Branching ratio of $B^+ \to \pi^+ K^0$ decay as a function of the mass of charged scalar $\Phi^-$. The blue band shows the experimental measurements, while the red curve represents the predicted branching ratio including new physics contributions and the horizontal purple line denotes the SM predictions at leading order of $\alpha_s$.}
\end{figure}
In Fig. \ref{fig:1}, we show the branching ratio of $B^+ \to \pi^+ K^0$ decay as a function of $m_{\Phi^-}$ with the coupling strength $\lambda$ fixed. It indicates that the charged scalars should be heavier than about $540$ GeV to be consistent with the charmless B decays. Noticed that the leading order SM prediction is about half less than the experimental measurements, as shown in Fig. \ref{fig:1}. This is because next-to-leading order amplitudes are not small in QCD factorization method, especially for chirally enhanced power corrections and annihilation diagrams (see, e.g.,  \cite{Beneke:2003zv,Du:2002cf,Li:2005wx,Cheng:2009cn}). But for the purpose of this paper, it should be enough to be confined at leading order.

The CDF dijet anomaly was explained in this flavor symmetry model by the process $\bar{u} s \to W^- \Phi_3^0$ via s-channel $\Phi_3^-$ exchange, with the cross section to be about $2$ pb. Now to satisfy B physics constraints, the charged scalar masses have to be raised from $220$ GeV to around $540$ GeV. As a result, the corresponding cross section must be well below $1$ pb, which is too small to account for the CDF dijet excess.
In addition, keeping $\lambda=1.4$ and $m_{\Phi^0}=160$ GeV unchanged while rasing the mass of charged scalars to $m_{\Phi^-}=540$ GeV, one might worry about its correction to the electroweak parameter \cite{Nelson:2011us}
\begin{align}\label{Eq:alphaT}
  \alpha T &=\frac{3}{32\pi^2 v^2} \left ( m_{\Phi^0}^2+m_{\Phi^-}^2-\frac{2m_{\Phi^-}^2 m_{\Phi^0}^2}{m_{\Phi^-}^2-m_{\Phi^0}^2}\log \frac{m_{\Phi^-}^2}{m_{\Phi^0}^2} \right )
\end{align}
which turns out to be $0.057$. Notice that $v=174$ GeV is taken in the above formula. Unfortunately this strongly contradicts the precision electroweak constraint \cite{Nakamura:2010zzi} $T=0.07(0.16) \pm 0.08$ assuming the Higgs mass $m_H=117(300)$ GeV. Therefore the mass splitting between the charged and the neutral scalars should be quite small to
satisfy the precision electroweak constraint, which means the neutral scalar masses should also be raised from $160$ GeV to around $500$ GeV.
For the color-singlet scalars in this mass range, one can see from Fig. 2 of \cite{Shu:2009xf} that the total forward-backward asymmetry of produced top pair may even be negative, which is opposite in sign to the experimental observations.

Another possibility is to reduce the coupling strength $\lambda$, and at the same time raising the charged scalar mass moderately to satisfy the constraints of electroweak parameter T and charmless B decays simultaneously. Noticed that taking the neutral scalar mass fixed at $160$ GeV, its correction to $\alpha T$ is $3.4 \times 10^{-3}$ with $m_{\Phi^-}=250$ GeV, which corresponds to $T=0.43$. This already deviates from the precision electroweak constraint $T=0.07(0.16) \pm 0.08$ by more than $3$ sigma. But if the standard model Higgs is very heavy, it will contribute negatively to the $T$ parameter. Keeping only leading logarithms in the Higgs mass, the contribution can be expressed approximately as \cite{Peskin:1991sw}
\begin{align}
             T \simeq -\frac{3}{16\pi \cos^2 \theta_W} \log \frac{m_H^2}{m_{H,ref}^2}~,
\end{align}
where $m_{H,ref}$ denotes the reference value of the Higgs mass. This means, for the SM Higgs as heavy as $1$ TeV, the precision electroweak constraint on T parameter would be around $T\simeq 0.40 \pm 0.08$, which is consistent with the flavor symmetric model with $m_{\Phi^-}=250$ GeV.
 
\begin{figure}
\includegraphics[scale=1.0]{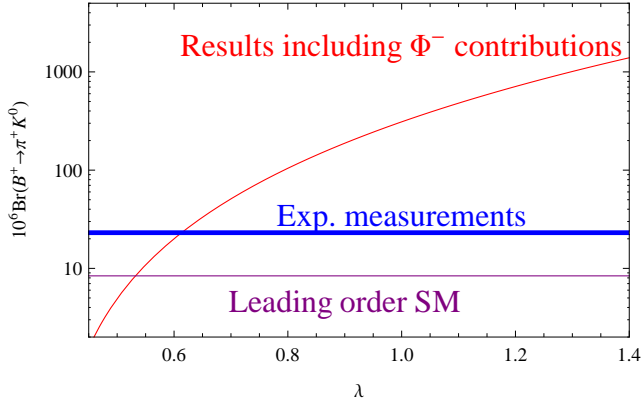} \caption{\label{fig:2} Branching ratio of $B^+ \to \pi^+ K^0$ decay as a function of the coupling strength $\lambda$, with the charged and neutral scalar masses taken at $250$ GeV and $160$ GeV, respectively. The meaning of the lines is the same as in Fig. \ref{fig:1}.}
\end{figure}

In any case, it is unlikely for the charged scalar mass in this model to be heavier than $250$ GeV. One may observe from Fig. \ref{fig:2} that, taking $m_{\Phi^-}=250$ GeV, $\lambda$ should be around $0.6$ to satisfy the restriction of $B^+ \to \pi^+ K^0$ decay. Notice that it was shown in \cite{Nelson:2011us} that in this model the top quark forward-backward asymmetry $A_{t\bar{t}}\simeq 0.13$ for $M_{t\bar{t}}> 450$ GeV with $\lambda=1.4$. It is then clear that this asymmetry must be well below $0.1$ if the coupling constant $\lambda$ is lowered to around $0.6$. Therefore it should be really hard, if not impossible, to explain the measured large forward-backward asymmetry of produced top pair under this circumstance.

As to the Wjj anomaly, the resonant production $\bar{u} s \to \Phi_3^-$ which subsequently decays to $W^- \Phi_3^0 \to W^- \bar{u} c$ can enhance the Wjj cross section. It is easy to calculate first the decay width of $\Phi_3^-$,
\begin{align}
\Gamma(\Phi_3^- \to \Phi_3^0 W^-)&=\frac{\alpha \lambda^{3/2}(m_{\Phi^-}^2,m_{\Phi^0}^2,m_W^2)}{8 \sin^2 \theta_W m_W^2 m_{\Phi^-}^3}
=0.18~\mbox{GeV}, \nonumber \\
\Gamma(\Phi_3^- \to \bar{u}s)&=\frac{N_c m_{\Phi^-}\lambda^2}{16\pi}=5.37 \left ( \frac{\lambda}{0.6} \right )^2 \mbox{GeV}~,
\end{align}
with the phase factor $\lambda(x,y,z)=x^2+y^2+z^2-2(xy+xz+yz)$. Correspondingly, the Wjj cross section is found to be $3.0$ pb for $\lambda=0.6$.
Actually, the Wjj cross section from resonant $\Phi_3^-$ production is not very sensitive to the value of $\lambda$, because the total width of $\Phi_3^-$ also changes with $\lambda$. For instance, the cross section is calculated to be $2.5$ pb with even smaller $\lambda=0.4$.

In summary, we consider the constraints of charmless B decays on a flavor symmetric scalar model proposed in \cite{Nelson:2011us}. The color-singlet weak-doublet scalars are introduced in the model which respects the flavor symmetry of $\left ( \prod_{i=1}^3 U(1)_{q_{Li}} \times U(1)_{u_{Ri}} \right ) \times U(3)_{d_R} \times \mathbb{Z}_3$. It was shown in \cite{Nelson:2011us} that the top quark forward-backward asymmetry and the dijet excess at CDF could be explained simultaneously with the parameters chosen as $\lambda=1.4$, $m_{\Phi^0}=160$ GeV and $m_{\Phi^-}=220$ GeV.
However the flavor symmetry of the scalars also contributes to $b \to s\bar{u}u$ decays. With the above fiducial values of the parameters, the pure penguin decay $B^+ \to \pi^+ K^0$ is predicted to have a branching ratio about one hundred times larger than the experimental results. To avoid this constraint, the charged scalars should be heavier than around $540$ GeV with $\lambda=1.4$ fixed. As a result, the production cross section of dijet plus a W boson would be too small to account for the CDF dijet excess. Furthermore, the precision electroweak constraints force the neutral scalar masses to be also around $500$ GeV. Then it also becomes hard for this model to account for the forward-backward asymmetry in top quark pair production.

Another possibility is to raise the charged scalar mass so that $\Phi^-_3 \to \Phi^0_3 W^-$ decay channel is allowed kinematically. In this scenario the Wjj cross section is enhanced due to the resonant production of $\Phi^-_3$ so that the coupling constant $\lambda$ may be lowered to evade the B physics constraint. Specifically, one may take $m_{\Phi^-}=250$ GeV, $m_{\Phi^0}=160$ GeV and $\lambda=0.6$. With this parameter set, the Wjj cross section is found to be $3$ pb, which is in the right range to explain the CDF dijet excess. But the scalar mass splitting will contribute to  $\alpha T=3.4 \times 10^{-3}$, which is about $3 \sigma$ deviation from the precision electroweak constraint. In addition, the smaller coupling strength will lead to too small $t\bar{t}$ forward-backward asymmetry to account for the experimental measurements.

\section*{Acknowledgement}
This work is supported in part by the National Science Foundation of China (No. 11075139 and No.10705024). G.Z is also supported in part by the Fundamental Research Funds for the Central Universities.

\end{document}